\def\deg{\ifmmode {^\circ}\else {$^\circ$}\fi}
\def\degree{\ifmmode {^\circ}\else {$^\circ$}\fi}
\def\mum{\ifmmode {\rm \,\mu {\rm m}}\else $\rm \,\mu {\rm m}$\fi}

\def\inch{\ifmmode ^{\prime \prime}\else $^{\prime \prime}$\fi}
\def\gs{\ifmmode {{\rm g~s^{-1}}}\else ${\rm g~s^{-1}}$\fi}
\def\msunyr{\ifmmode {M_{\odot}~{\rm yr^{-1}}}\else $M_{\odot}~{\rm yr^{-1}}$\fi}
\def\msun{\ifmmode {M_{\odot}}\else $M_{\odot}$\fi}
\def\rsun{\ifmmode {R_{\odot}}\else $R_{\odot}$\fi}
\def\lsun{\ifmmode {L_{\odot}}\else $L_{\odot}$\fi}
\def\mstar{\ifmmode {M_{\star}}\else $M_{\star}$\fi}
\def\rstar{\ifmmode {R_{\star}}\else $R_{\star}$\fi}
\def\tstar{\ifmmode {T_{\star}}\else $T_{\star}$\fi}
\def\lstar{\ifmmode {L_{\star}}\else $L_{\star}$\fi}
\def\mwd{\ifmmode {M_{wd}}\else $M_{wd}$\fi}
\def\rwd{\ifmmode {R_{wd}}\else $R_{wd}$\fi}
\def\twd{\ifmmode {T_{wd}}\else $T_{wd}$\fi}
\def\lwd{\ifmmode {L_{wd}}\else $L_{wd}$\fi}
\def\md{\ifmmode {M_d}\else $M_d$\fi}
\def\ld{\ifmmode {L_d}\else $L_d$\fi}
\def\ad{\ifmmode A_d\else $A_d$\fi}
\def\ldlwd{\ifmmode L_d / L_{wd}\else $L_d / L_{wd}$\fi}
\def\ldlstar{\ifmmode L_d / L_\star\else $L_d / L_{\star}$\fi}
\def\rearth{\ifmmode {\rm R_{\oplus}}\else $\rm R_{\oplus}$\fi}
\def\mearth{\ifmmode {\rm M_{\oplus}}\else $\rm M_{\oplus}$\fi}
\def\qc{\ifmmode Q_c\else $Q_c$\fi}
\def\qdstar{\ifmmode Q_D^\star\else $Q_D^\star$\fi}
\def\rt{\ifmmode r_t\else $r_t$\fi}
\def\vc{\ifmmode v_c\else $v_c$\fi}
\def\vsqd{\ifmmode v^2 / Q_D^\star\else $v^2 / Q_D^\star$\fi}
\def\kms{\ifmmode {\rm km~s^{-1}}\else $\rm km~s^{-1}$\fi}
\def\ms{\ifmmode {\rm m~s^{-1}}\else $\rm m~s^{-1}$\fi}
\def\vrel{\ifmmode v_{rel}\else $v_{rel}$\fi}
\def\mdot{\ifmmode \dot{M}\else $\dot{M}$\fi}
\def\mdotz{\ifmmode \dot{M}_0\else $\dot{M}_0$\fi}
\def\mesc{\ifmmode m_{esc}\else $m_{esc}$\fi}
\def\rmin{\ifmmode r_{min}\else $r_{min}$\fi}
\def\rmax{\ifmmode r_{max}\else $r_{max}$\fi}
\def\xmax{\ifmmode x_{max}\else $x_{max}$\fi}
\def\mmin{\ifmmode m_{min}\else $m_{min}$\fi}
\def\mmax{\ifmmode m_{max}\else $m_{max}$\fi}
\def\rmind{\ifmmode r_{min,d}\else $r_{min,d}$\fi}
\def\rmaxd{\ifmmode r_{max,d}\else $r_{max,d}$\fi}
\def\mmaxd{\ifmmode m_{max,d}\else $m_{max,d}$\fi}
\def\vrad{\ifmmode v_{rad}\else $v_{rad}$\fi}
\def\qz{\ifmmode q_{0}\else $q_{0}$\fi}
\def\qi{\ifmmode q_{i}\else $q_{i}$\fi}
\def\ql{\ifmmode q_{l}\else $q_{l}$\fi}
\def\qs{\ifmmode q_{s}\else $q_{s}$\fi}
\def\vhill{\ifmmode v_H\else $r_H$\fi}
\def\rhill{\ifmmode r_H\else $r_H$\fi}
\def\Rhill{\ifmmode R_H\else $R_H$\fi}
\def\rbrk{\ifmmode r_{brk}\else $r_{brk}$\fi}
\def\rdamp{\ifmmode r_{damp}\else $r_{damp}$\fi}
\def\rin{\ifmmode r_{in}\else $r_{in}$\fi}
\def\rout{\ifmmode r_{out}\else $r_{out}$\fi}
\def\tin{\ifmmode t_{in}\else $t_{in}$\fi}
\def\tout{\ifmmode t_{out}\else $t_{out}$\fi}
\def\ain{\ifmmode a_{in}\else $a_{in}$\fi}
\def\aout{\ifmmode a_{out}\else $a_{out}$\fi}
\def\r0{\ifmmode r_{0}\else $r_{0}$\fi}
\def\R0{\ifmmode R_{0}\else $R_{0}$\fi}
\def\m0{\ifmmode m_{0}\else $m_{0}$\fi}
\def\mone{\ifmmode m_{1}\else $m_{1}$\fi}
\def\mtwo{\ifmmode m_{2}\else $m_{2}$\fi}
\def\atwo{\ifmmode a_{2}\else $a_{2}$\fi}
\def\etwo{\ifmmode e_{2}\else $e_{2}$\fi}
\def\mf{\ifmmode m_{f}\else $m_{f}$\fi}
\def\af{\ifmmode a_{f}\else $a_{f}$\fi}
\def\ef{\ifmmode e_{f}\else $e_{f}$\fi}
\def\M0{\ifmmode M_{0}\else $M_{0}$\fi}
\def\amax{\ifmmode a_{max}\else $a_{max}$\fi}
\def\a0{\ifmmode a_{0}\else $a_{0}$\fi}
\def\e0{\ifmmode e_{0}\else $e_{0}$\fi}
\def\v0{\ifmmode v_{0}\else $v_{0}$\fi}
\def\xm{\ifmmode x_{m}\else $x_{m}$\fi}
\def\sigz{\ifmmode \Sigma_0\else $\Sigma_0$\fi}
\def\ergg{\ifmmode {\rm erg~g^{-1}}\else ${\rm erg~g^{-1}}$\fi}
\def\ergs{\ifmmode {\rm erg~s^{-1}}\else ${\rm erg~s^{-1}}$\fi}
\def\gyr{\ifmmode {\rm g~yr^{-1}}\else ${\rm g~yr^{-1}}$\fi}
\def\cms{\ifmmode {\rm cm~s^{-1}}\else ${\rm cm~s^{-1}}$\fi}
\def\gcms{\ifmmode {\rm g~cm^{-2}}\else $\rm g~cm^{-2}$\fi}
\def\gcmc{\ifmmode {\rm g~cm^{-3}}\else $\rm g~cm^{-3}$\fi}
\def\atil{\ifmmode {\tilde{a}}\else $\tilde{a}$\fi}
\def\ttil{\ifmmode {\tilde{t}}\else $\tilde{t}$\fi}
\def\sqrttt{\ifmmode {\tilde{t}^{1/2}}\else $\tilde{t}^{1/2}$\fi}

\def\herschel{{\it Herschel}}
\def\spitz{{\it Spitzer}}

\def\wise{{\it WISE}}

\def\mp{\ifmmode m_P\else $m_P$\fi}
\def\mc{\ifmmode m_C\else $m_C$\fi}
\def\mh{\ifmmode m_H\else $m_H$\fi}
\def\mk{\ifmmode m_K\else $m_K$\fi}
\def\ms{\ifmmode m_S\else $m_S$\fi}
\def\mn{\ifmmode m_N\else $m_N$\fi}
\def\rp{\ifmmode r_P\else $r_P$\fi}
\def\rc{\ifmmode r_C\else $r_C$\fi}
\def\apc{\ifmmode a_{PC}\else $a_{PC}$\fi}
\def\mpc{\ifmmode m_{PC}\else $m_{PC}$\fi}
\def\epc{\ifmmode e_{PC}\else $e_{PC}$\fi}

\documentclass[]{aastex631}

\shorttitle{Takeout \& Delivery}
\shortauthors{Najita \& Kenyon}
\graphicspath{{./}{figures/}}

\begin{document}

\title{Takeout and Delivery: 
Erasing the Dusty Signature of
Late-stage Terrestrial Planet Formation}

\correspondingauthor{Joan R. Najita}
\email{joan.najita@noirlab.edu, skenyon@cfa.harvard.edu}

\author[0000-0002-5758-150X]{Joan R. Najita}
\affiliation{NSF's NOIRLab, 950 N. Cherry Avenue, Tucson, AZ 85719, USA}

\author[0000-0003-0214-609X]{Scott J. Kenyon}
\affiliation{Smithsonian Astrophysical Observatory, 60 Garden Street, 
Cambridge, MA 02138, USA}

\begin{abstract}

The formation of planets like Earth is expected to conclude with a series of late-stage giant impacts that generate warm dusty debris, the most anticipated visible signpost of terrestrial planet formation in progress. 
While there is now evidence that Earth-sized terrestrial planets orbit a significant fraction of solar-type stars, the anticipated dusty debris signature of their formation is rarely detected. 
Here we discuss several
ways in which our current ideas about terrestrial planet formation imply transport mechanisms capable of erasing the anticipated debris signature.
A tenuous gas disk may be regenerated via ``takeout'' (i.e., the liberation of planetary atmospheres in giant impacts) or ``delivery'' (i.e., by asteroids and comets flung into the terrestrial planet region) at a level sufficient to remove the warm debris. 
The powerful stellar wind from a young star can also act, its delivered wind momentum producing a drag that removes warm debris. 
If such processes are efficient, 
terrestrial planets may assemble inconspicuously, with little publicity and hoopla accompanying their birth. 
Alternatively, the rarity of warm excesses may imply that terrestrial planets typically form very early, emerging fully formed from the nebular phase without undergoing late-stage giant impacts. 
In either case, the observable signposts of terrestrial planet formation appear more challenging to detect than previously assumed.
We discuss observational tests of these ideas. 

\end{abstract}


\section{Introduction} \label{sec:intro}

The ``classical'' picture of terrestrial planet formation begins with the formation of km-sized or larger planetesimals within a primordial gaseous disk around a solar-type star \citep[e.g.,][]{liss1987,weth1993,weiden1997b,chambers1998,gold2004,kb2006,raymond2014a}. The planetesimals collide and merge into lunar- to Mars-sized protoplanets, which interact gravitationally once the gaseous disk dissipates, leading to collisions, mergers, and ever-larger protoplanets 
\citep[e.g.,][]{iwasaki2001,iwasaki2002,kominami2002,kominami2004}.
Over the first 50--100~Myr,``giant impacts'' among the few remaining protoplanets---as in the moon-forming collision at $\sim$ 30--50~Myr in the Solar System \citep[e.g.,][and references therein]{canup2021}---build the final set of terrestrial planets \citep[e.g.,][]{lammer2021}. Throughout this phase, 
giant impacts and collisions among remaining planetesimals 
are expected to 
generate significant debris in the form of cm-sized and smaller particles. The infrared emission from this warm debris, 
{which is expected to be bright and readily detectable,}
is the most anticipated visible signpost of ongoing terrestrial planet formation
\citep[e.g.,][]{kb2002b,kb2004b,raymond2011,raymond2012,genda2015a}. 

In \citet{knb2016}, we described how this picture leads to a current conundrum: although planets like Earth and Venus are believed to occur commonly around mature solar-type stars based on estimates from transit surveys
\citep[e.g., $\sim 20$\% occurrence rate;][]{petigura2013b,burke2015,bryson2021}, the warm excesses 
expected to accompany their formation 
\citep[e.g.,][]{kb2004b,genda2015a,kb2016a,kobayashi2019} are detected only rarely, in $\le 2$\%--3\% of young solar-type stars.
The discrepancy can be resolved if terrestrial planets are not as common as currently estimated.  
Alternatively, terrestrial planet formation may work 
differently than we think, with planets completing their formation on much shorter timescales than in our current picture, while they are embedded in the primordial gaseous disk.
A more prosaic possibility is that there are unexamined processes that remove the dusty debris signature of late-stage terrestrial planet formation. 

Here we explore two possible dust removal mechanisms: gas drag and stellar wind drag.
As described in \citet{knb2016}, if the inner regions of planetary systems evolve so that 
they end up possessing tenuous gas reservoirs at the epoch of final terrestrial planet assembly, at a level of $\sim 10^{-5} - 10^{-6}$ of the minimum mass Solar nebula (MMSN),
the gas can remove the dusty debris associated with terrestrial planet formation through a combination of aerodynamic drag and radiation pressure \citep[see also][]{take2001,krumholz2020}.
A $10^{-5}$ MMSN column density of 0.02\,\gcms\ within 2\,au corresponds to a gas mass of 0.01\,\mearth\ (equivalent to $\sim$ 43 Earth oceans). 



Several possible pathways to such tenuous gas disks emerge naturally from our current picture of terrestrial planet formation.
Gaseous planetary atmospheres, primordial or secondary, might be liberated through 
giant impacts, 
the tragic loss of planetary atmospheres a boon to the removal of the planet's formation signature.  
Alternatively, gas may be brought in by evaporating icy comets and asteroids that have been scattered into the terrestrial planet region. 
%
%
Stellar wind drag is another possible dust removal mechanism, in which dust grains in the terrestrial planet region 
are dragged into the
central star through interactions with a high velocity stellar wind
\citep[e.g.,][]{burns1979,plavchan2005,klacka2013,spalding2020}. While 
this mechanism is unimportant for dust around solar-type stars at Gyr
ages 
and is weaker at large orbital distances, it can be a much more potent dust removal mechanism 
at the small orbital radii associated with terrestrial planet formation 
and at the 
enhanced wind mass loss rates thought to characterize young stars 
(ages $\lesssim 100$\,Myr). 

In Section 2, we recap the predicted dusty signature that is predicted to accompany late terrestrial planet formation and the current observational limits.
We then describe the gas reservoirs that are expected to be liberated from planetary atmospheres by giant impacts (Section 3) and delivered to inner solar systems by asteroids and comets (Section 4). Section 5 describes the possible role of stellar wind drag in removing collisional debris from the terrestrial planet region. These ideas and future directions are discussed in Section 6 and summarized in Section 7. 

\section{Predicted IR Excesses and Observational Limits} 



To establish the conditions required for a detectable IR 
excess from solid material at 1~au, we follow \citet{knb2016}.  
{\it Spitzer} observations at 16--24~\mum\ (8--12~\mum) are sufficient 
to detect solid material at $a$ = 1~au with a temperature $T_d \sim$ 300~K 
and a fractional luminosity $\ldlstar\ = A_d / 4 \pi a^2 \gtrsim 10^{-4}$
($\gtrsim 3 \times 10^{-4}$), where $L_d$ is the thermal luminosity of the 
solids, \lstar\ is the luminosity of the central star, and $A_d$ is the 
cross-sectional area of the solids. For a given sensitivity at 16--24~\mum, 
hotter (colder) dust is harder (easier) to detect \citep[see Figure 1 
of][]{knb2016}. 

{
Although 24~\mum\ excesses are fairly common among nearby solar-type stars
with ages of 10--100~Myr, 8--16~\mum\ excesses are rare \citep[e.g.,][]
{stauffer2005,silver2006,currie2007b,rhee2008,dahm2009,carp2009a,carp2009b,melis2010,stauffer2010,
chen2011,beich2011,smith2011,zuck2011,luhman2012b,ribas2012,urban2012,zuck2012,
jackson2012,kw2012,kenn2013a,ballering2013,clout2014,matthews2014,vican2014,
patel2014,ishihara2017,melis2021,moor2021}. 
As discussed in \citet{knb2016}, the 
great majority of 24~\mum\ excess sources lack excess emission at shorter wavelengths, implying that they arise from
cold material
with a dust temperature $T_d \lesssim$ 150~K. The high (low) frequency of cold
(warm) dust from short wavelength \spitz\ and \wise\ studies is consistent 
with the $\sim$ 25\% frequency of cold excesses among solar-type stars derived
from other \spitz\ and \herschel\ programs \citep[e.g.,][and references therein]
{bryden2009,eiroa2013,sibthorpe2018,matra2018,nkb2022}. Thus, we conclude that
the warm excesses associated with solar-type stars with ages $\sim$ 10--100~Myr 
are rare.
}

Converting an 
observational limit on 
fractional luminosity
into an estimate of the 
total mass of solids requires an adopted size distribution as a function 
of particle radius $r$, $N(r)$. 
{In a simple 
system, where all solids have the same binding energy and the 
collision velocity is just large enough to catastrophically 
fragment particles, $N(r) \propto r^{-3.5}$ \citep{kb2020}.} 
The mass is then
$M_d = (4/3) \rho A_d (r_{min} r_{max})^{1/2}$, where $\rho$ 
is the mass
density of a particle and \rmax\ is the radius of the largest particle. 
Setting \rmax\ = 300~km, $\rho$ = 3~\gcmc, and $\ldlstar = 10^{-4}$, the 
ensemble of solids has a total cross-sectional area 
$A_d \approx 3 \times 10^{23}\,{\rm cm}^2$ and a total mass
$M_d \approx 5.5 \times 10^{25}$~g $\approx$ 0.7 lunar masses.

{
Systems with a size-dependent binding energy have 
wavy size distributions that are somewhat steeper than 
$N(r) \propto r^{-3.5}$ \citep[e.g.,][]{campo1994a,obrien2003,
koba2010a,wyatt2011}. Adopting the bulk properties of basalt \citep{benz1999}, 
a collision velocity large enough to catastrophically fragment 300~km 
planetesimals, and a collisional cascade in equilibrium from 300~km 
to 1~\mum\ \citep[e.g.,][]{wyatt2011,kb2020}, the mass is $\sim$ 65\% 
smaller than estimated above: $M_d \approx 3.7 \times 10^{25}$~g (0.5 
lunar masses). Although different choices for the bulk properties of solids
and the collision velocity result in somewhat different masses,
0.2--1~lunar masses in solids is sufficient to generate the cross-sectional
area, $A_d \approx 3 \times 10^{23}\,{\rm cm}^2$, required for material with
\ldlstar $\approx 10^{-4}$ at $a \approx$ 1~au.}

These estimates are consistent with solid mass and dust luminosity 
estimates derived in numerical calculations of the giant impact phase 
\citep[e.g.,][]{kb2004b,jackson2012,genda2015a,kb2016a,kobayashi2019}. 
In these simulations, collisions of lunar-mass to Mars-mass protoplanets 
usually generate larger merged objects and ejected fragments with radii of 
100--1000~km and total masses of $\lesssim$ 1--2 lunar masses. As protoplanets
grow into larger and larger planets, the fragments generate a collisional 
cascade with a detectable 
{infrared excess at 16--24~\mum\ and}
a dust luminosity, 
\ldlstar\ $\gtrsim 10^{-4}$,
throughout the 10--100~Myr phase
of rocky planet building. 

During the collisional cascade, high velocity collisions gradually grind 
down larger solids into smaller solids, which are ejected from the system 
by stellar radiation pressure \citep[e.g.,][]{wyatt2008,koba2014,
matthews2014,kb2016a,kobayashi2019,marino2022}.
In the standard kinetic model, the evolution time for the cascade is 
roughly equivalent to the time required for a collision between the two 
largest particles in the swarm, which depends on the distance from the 
central star, the radius \rmax\ of the largest solid involved in the 
cascade, and the mass in solids with radii $r \le \rmax$ 
\citep[e.g.,][]{wyatt2002,dom2003,wyatt2008,krivov2008,kb2017a,
krivov2021}. If the solids reside in an annulus with a radial width 
$\delta a$ that is larger than the radial excursion of a typical solid, 
$\delta a \gg e a$, where $e$ is the orbital eccentricity, the time 
scale to remove the largest particles in the cascade is 
$t_d = \rmax\ \rho P \alpha / 12 \pi \Sigma_d$, where 
$P$ is the orbital period and $\Sigma_d$ is the surface 
density of solids at a distance $a$ from the host star \citep[e.g.,][and 
references therein]{kb2017a}. 
{In this expression, the $\alpha$ term encodes the
dependence of the time scale on the collision velocity and the binding
energy of the largest solids (for other approaches, see \citet{bottke2005},
\citet{koba2010a}, and \citet{kw2010}). Defining $v$ as the center-of-mass
collision velocity and \qdstar\ as the collision energy required to remove
half the mass of a pair of colliding planetesimals and disperse it to
infinity, $\alpha$ is a function of the ratio \vsqd\ \citep{kb2017a}.
When high velocity collisions just shatter the largest objects
(\vsqd\ $\approx$ 8), $\alpha \approx 5$. Factor of three larger velocities
yields $\alpha \approx 0.7$ \citep[see Fig.~2 of][]{kb2017a}.
}

Adopting  $\Sigma_d \approx M_d / \pi a^2$, the lifetime of the cascade is
\begin{equation}
\label{eq:t0}
t_d \approx {\rm 30~Myr}~\alpha
\left ( \frac{a}{\rm 1~au} \right )^{3/2} 
\left ( \frac{\rmax}{\rm 300~km} \right )^{1/2}
\left ( \frac{1~\mum}{\rmin} \right )^{1/2}
\left ( \frac{10^{-4}}{\ldlstar} \right ) ~ .
\end{equation} 
As a result, the debris produced by giant impacts is expected to remain bright 
and observable for an extended period of time, in the absence of removal 
mechanisms other than collisions and stellar radiation pressure.
{
When $\alpha \approx$ 0.5--3, the collision time in eq.~\ref{eq:t0} is
similar to the lifetime of detectable warm dust in numerical calculations
\citep{kb2004b,raymond2011,genda2015a,kb2016a}. In Fig.~6 of \citet{genda2015a},
the 24~\mum\ flux from the debris of giant impacts equals or exceeds the
flux from the central star (and the detection limit for \spitz\ at 16-24~\mum) 
for $\sim$ 100~Myr.  
Although the bright pulse of 
emission generated in a single giant impact
lasts for only a few Myr, 
the broad pedestal of excess emission produced by
dust 
from the ensemble of giant impacts persists
for $\gtrsim$ 100~Myr. Similarly, calculations described in \citet{kb2004b,kb2016a}
have dust luminosities \ldlstar\ $\gtrsim 10^{-4}$ for evolution
times $\lesssim$ 100~Myr. 
The rarity of infrared excess emission detections at short wavelengths (e.g., at 16~\mum) demonstrates that such warm debris is rare.}

\section{Gas from Liberated Planetary Atmospheres}
\label{sec:atmospheres}
\subsection{Primordial Atmospheres} \label{sec:primordial}

Solar System formation models suggest that terrestrial planet formation 
was well underway during the protosolar nebular phase. 
Planets like Earth and Venus may have acquired a significant fraction of their mass (perhaps $> 50$\%) prior to the dissipation of the nebula
\citep{walsh2011a,obrien2014,raymond2014a,lammer2021}, a perspective that
is supported by some interpretations of meteoritic data \citep{boyet2005,yu2011}. 
Terrestrial planets that acquire a significant mass 
while the nebular gas disk is present 
are expected to capture tenuous primordial atmospheres whose mass depends on the nebular density \citep[e.g.,][]{rafikov2006}. 
A protoplanet with mass $M_p=0.55\,\mearth$ could acquire an atmosphere that is $M_{\rm atm}/M_p = 1$\% of the planet's mass \citep{lammer2020a}. A lower-mass protoplanet with $M_p \approx$ 0.1--0.3\,\mearth\ may acquire an atmosphere
that is 0.01\% to 1\% of the planet's mass \citep{ginzburg2016,lammer2020a}. These atmospheres 
are similar to the 0.001--0.01 \mearth\ needed for aerodynamic drag to remove dusty debris from the terrestrial planet region (Section 1).

As the nebula dissipates, the planetary atmosphere
adjusts to the declining external pressure from the nebula and the
increasing exposure to direct irradiation by the host star.
Planets orbiting close to their stars are more likely than planets at larger distances to lose some or all of their primordial atmosphere before the giant impact phase commences. 

Theory 
suggests that low mass planets ($\lesssim$ 5--10~\mearth) orbiting 
within $\sim 0.5$~au lose a significant fraction, perhaps all, of their 
atmospheres through photoevaporation \citep{owen2013,lopez2013,jin2014,
owen2016c,owen2017b,jin2018,mordasini2020,venturini2020} or core-powered 
mass loss \citep{ginzburg2016,ginzburg2018,gupta2019}. 
EUV/X-ray photoevaporation is particularly efficient in reducing planetary atmospheres at orbital radii $\lesssim 0.2$\,au, with the transformation occurring on a timescale $\lesssim$ 100 Myr \citep{owen2013}. 
Core-powered mass loss operates over longer 
($\sim$ Gyr) timescales
\citep{ginzburg2018}.
Planetesimal impacts remove atmospheres on even longer timescales. 
Among planets with orbital periods $\lesssim$ 30~days, observations 
provide strong support for the `radius' or `evaporation' valley predicted 
from these studies\footnote{For an alternative approach to the radius 
valley, see \citet{elee2021}. 
}
\citep[e.g.,][]{fulton2017,vaneylen2018,
fulton2018,hardegree2020,petigura2020}. Better statistics for the 
planetary mass--radius relation at 0.2--0.6~au would test these theories in greater detail \citep[e.g.,][]{rogers2021}.


Beyond 0.5--0.6~au, planets have a better chance of retaining their 
primordial atmospheres. At these larger distances, photoevaporation and 
core-driven mass loss are less effective \citep[e.g.,][]{mordasini2020}. 
Planetesimal impacts---which may remove or add material to the planet's 
atmosphere---then become more important \citep{wyatt2020,kegerreis2020a,
kegerreis2020b}. Although the relative importance of these processes among
exoplanets is uncertain, abundance analyses of Venus and Earth provide 
some clues. 
Detailed models for the 
atmospheric evolution of Earth (Venus) conclude that protoplanets with
masses exceeding $\sim$ 0.5--0.6~\mearth\ ($\sim$ 0.85--0.9~\mearth) 
retain their primordial atmospheres following the dissipation of the disk,
while those with lower masses do not \citep[e.g.,][]{lammer2020a}.

\subsection{Secondary Atmospheres} \label{sec:secondary}
Whether they retain any portion of their primordial atmospheres, terrestrial planets can also generate secondary atmospheres through outgassing. Recently formed terrestrial planets are expected to possess significant magma oceans, involving part or all of the silicate mantle
\citep[e.g.,][]{urey1955,elkins2008a,elkins2011,elkins2012,lammer2018}. Melting is induced by heating from multiple processes: the decay of short-lived radiogenic isotopes \citep[e.g., $^{26}$Al;][]{neumann2014}, the conversion of gravitational potential energy into heat when the core and mantle decouple \citep[e.g.,][]{rubie2015}, and kinetic energy from protoplanet impacts \citep[e.g.,][]{tonks1993,
elkins2004,monteux2014,manske2021}. As magma oceans solidify, they are  expected to expel copious amounts of gas. 
Some portion of the gaseous atmosphere is then expected to be ejected in the giant impacts that build the final masses of terrestrial planets.

Estimates from the literature aimed at 
explaining the known properties of Solar System planets predict that a planet can outgas
70\%--100\% of its initial carbon and water into the atmosphere 
\citep[][and references therein]{elkins2008a,hamano2013,erkaev2014,lichtenegger2016,
odert2016,massol2016,lammer2018}.
%
We can consider in a more general way the volatiles available to terrestrial planets as they are assembled, independent of how the assembly works in detail.
The volatile budget of the solids from which planets form establishes an upper envelope for the reservoir available for a regenerated gaseous atmosphere.
The water abundance of the bulk silicate Earth, 0.6\%--3.6\% by 
weight \citep[e.g.,][]{daly2021},
lies between that of carbonaceous chondrites \citep[2\%--13\% by weight;]
[]{alexander2012,marty2016,mccubbin2019,shimizu2021,alexander2022} and that of enstatite and ordinary chondrites 
\citep[0.1\%--0.5\% by weight; ][]{piani2018a,piani2018b,piani2020,
bates2021,king2021}.


Taking a conservative approach, if Earth formed from material that was 1\%
volatile by weight, 
and the volatiles are completely outgassed, as in some of the detailed models above, 
the mass of the secondary atmosphere would be similar to the gas reservoir needed
to remove the debris signature (Section 1).
If carbonaceous chondrite-like materials contribute significantly to the planet's mass budget,
the volatile budget could be significantly larger. Given the likely importance of inward pebble drift and pebble accretion in building terrestrial planet cores \citep{lammer2020a,lammer2020b}, it seems plausible that volatile-rich carbonaceous chondrites could contribute a significant fraction of a terrestrial planet's mass \citep{alexander2022}.


\begin{figure}[t]
\begin{center}
\includegraphics[width=5.5in]{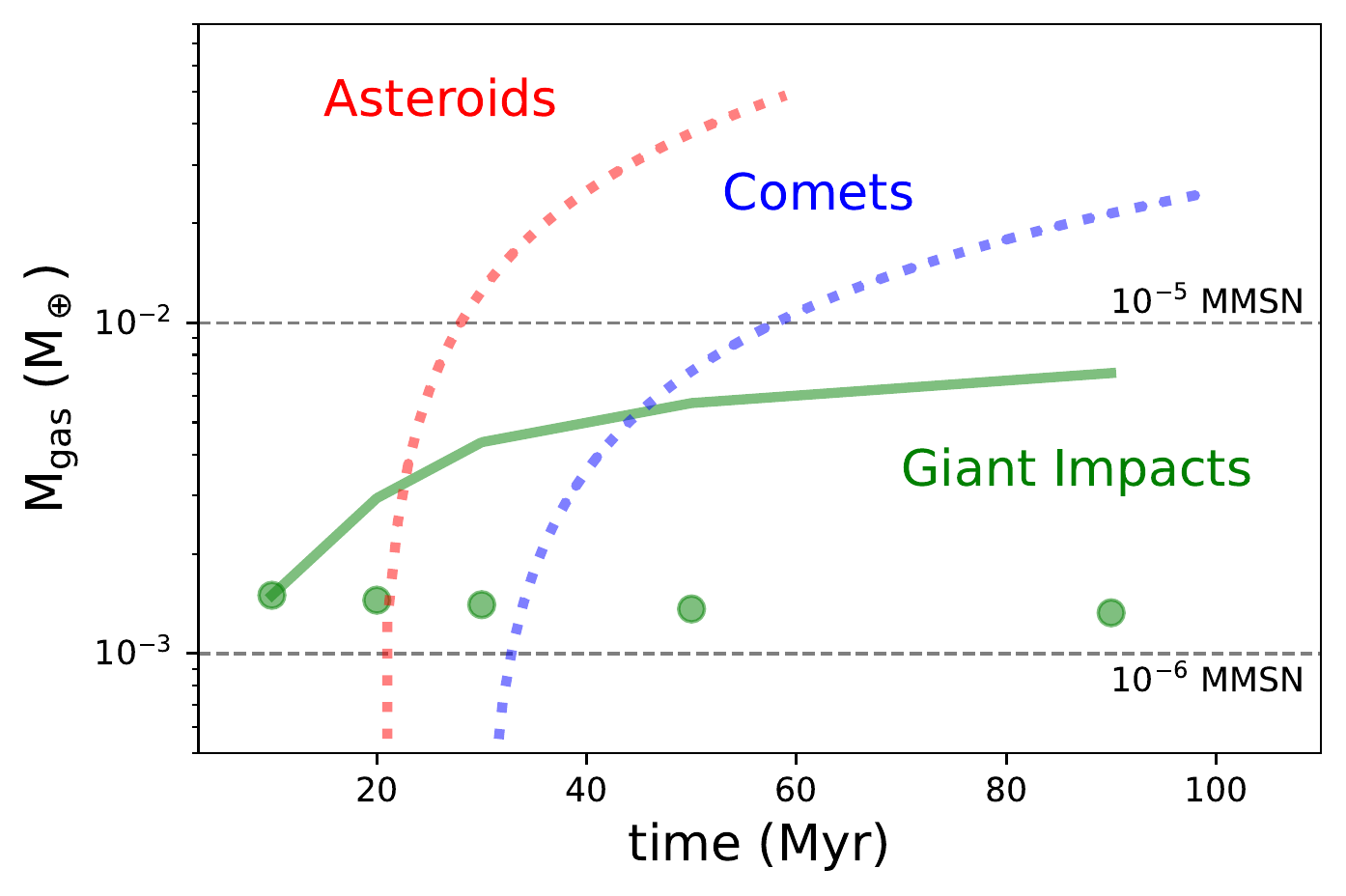}
\vskip -2ex
\caption{
Schematic illustration of possible contributors to the regeneration of a tenuous gas disk: gaseous atmospheres liberated by a sequence of giant impacts (green dots) and their cumulative contribution (green line), 
{as described in Section 3,} and the cumulative mass in volatiles delivered by scattered comets (dashed blue line) and asteroids (dashed red line),
{as described in Section 4}. 
{For illustration purposes,} comets and asteroids are assumed to deliver gas at constant mass input rates over the time intervals shown. 
}
\label{fig:mgas}
\end{center}
\end{figure}

\subsection{Giant Impacts}
Collisions between protoplanets and planets will liberate some or all of
the outgassed atmospheres
\citep[{primordial and/or secondary}; e.g.,][]{ahrens1993,genda2003b,schlichting2015,inamdar2015,
yalinewich2019,kegerreis2020a,kegerreis2020b,biersteker2019,biersteker2021},
regenerating a gaseous disk.
As a concrete example, we imagine starting with a 0.5\,\mearth\ planet that experiences 5 successive giant impacts with protoplanets, each of which has a mass 0.1\,\mearth. For simplicity, we assume 
{no primordial atmosphere and} that the volatiles in the planet and protoplanets have been completely outgassed into their atmospheres.
Following the discussion in \citet{inamdar2015}, we assume that each impact removes the atmosphere of the impactor completely. The fraction of the planet’s atmosphere that is removed depends on the quantity $v_{\rm imp} m_{\rm imp} / v_{\rm esc} M_{\rm core}$,
where $v_{\rm imp}$ and $m_{\rm imp}$
are the velocity and mass of the impactor,
$v_{\rm esc}$ is the planet's escape velocity, and
$M_{\rm core}$ is the planet's mass.
Assuming $v_{\rm imp}/v_{\rm esc} \simeq 1$ and $m_{\rm imp}/M_{\rm core} \simeq 0.2,$ the fraction of the planet’s atmosphere that is removed is $\sim 0.1$
as long as the mass of the planet's atmosphere is much less than the mass of its core
\citep{inamdar2015}.



The successive collisions build up a tenuous gaseous disk.
If all of the colliding bodies have a volatile fraction of 
1\% by mass, 
each giant impact contributes 
$10^{-3}\,\mearth$ 
of gas from the impactor and
$5 \times 10^{-4}\,\mearth$ to $3 \times 10^{-4}\,\mearth$
of gas lost from the planet's atmosphere; the amount declines with successive impacts as the planet's atmosphere is depleted.
Each impact ejects 
$\sim 1.4 \times 10^{-3}\,\mearth$ of gas from the planet and impactor (Figure 1, green points),
and the total gas mass liberated 
after 5 impacts is
$\sim 7\times 10^{-3}~\mearth$
(Figure~\ref{fig:mgas}, green line).
%
By regenerating a gaseous inner disk through outgassing and giant impacts, the gas is available exactly when needed to remove the dust signature generated by the same giant impacts.  

The liberated gas is comparable to the mass needed to remove the expected debris signature \citep{knb2016}.
{
To estimate the removal time, we assume the gas extends from 
$a_{in} \lesssim$ 0.5~au to $a_{out} \gtrsim$ 2--3~au, orbits
the central star at the local circular velocity, and has a
radial surface density distribution $\Sigma \propto a^{-1}$.
The central star has a mass \mstar\ = 1~\msun\ and luminosity
\lstar\ = 1~\lsun.
As in \citet{knb2016}, we combine the approaches of 
\citet{weiden1977a} and \citet{take2001} and solve for the 
radial and azimuthal velocity of particles relative to the gas.
Our approach assumes a vertical scale height 
$H/a = 0.03 ~ (a / \rm{1~au})^{1/8}$, a gas temperature 
$T_g = 278 ~ (a / {\rm 1~au})^{1/2}$, and other parameters 
appropriate for a gaseous disk \citep[see also][]
{ada1976,kh1987,chiang1997,take2001,raf2004,chiang2010,
armitage2013,youdin2013}. Additional details are available in Appendix A.2 of \citet{knb2016}. Other choices for the 
exponents in the relations for $\Sigma$, $T_g$, and $H$ would have a modest impact on the results.
}

\begin{figure}[t]
\begin{center}
\includegraphics[width=5.0in]{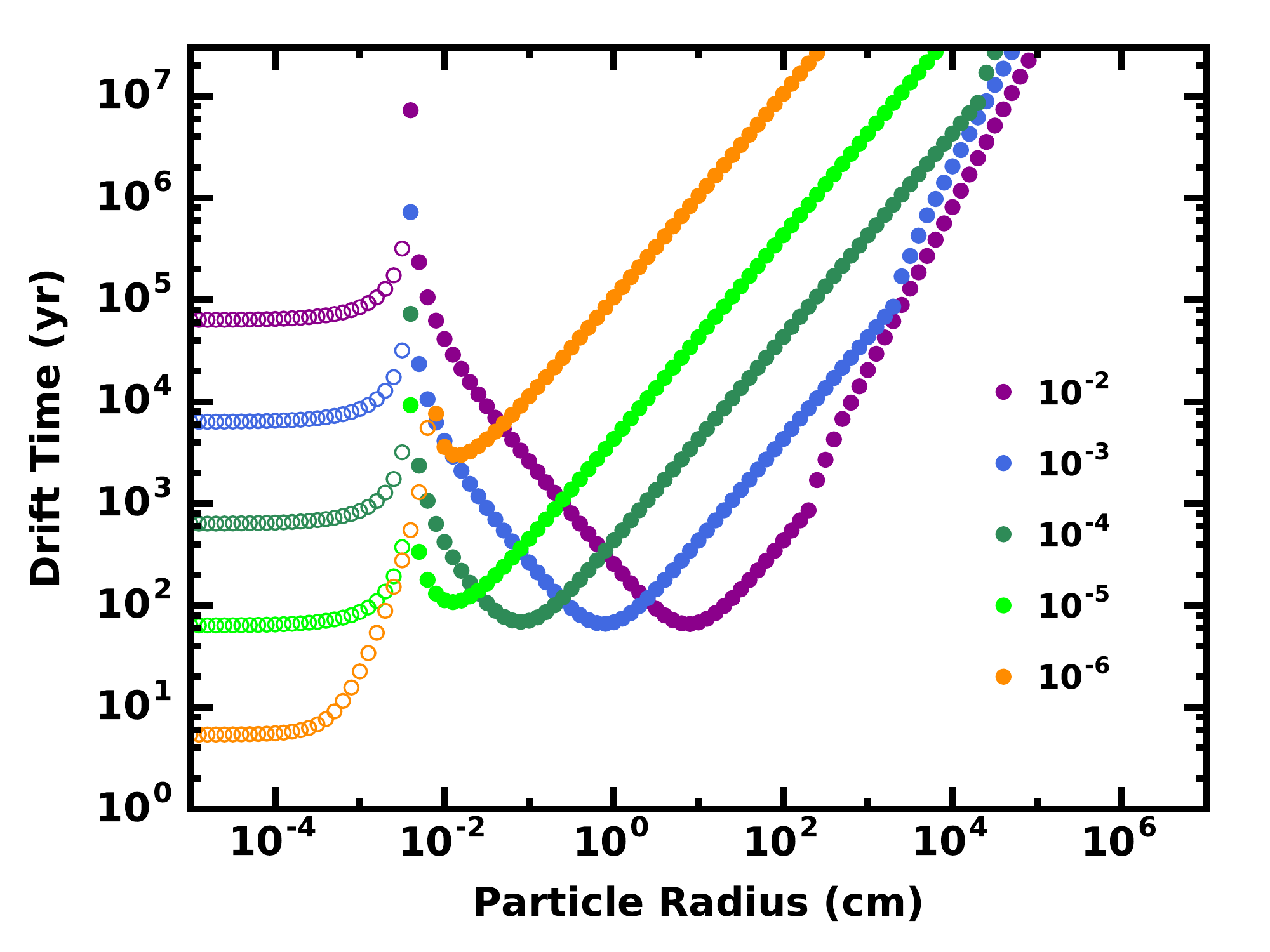}
\vskip -2ex
\caption{\label{fig:gasdrag}
Radial drift time scales for solid particles within a gaseous disk at 
1~au. The legend lists the disk surface density relative to a MMSN with
$\Sigma_0$ = 2000~\gcmc\ at 1~au. Filled (open) circles indicate inward 
(outward) drift. For this configuration, inward drift from gas drag 
balances outward drift from radiation for particles with $r \approx$ 
40~\mum. As the gas surface density at 1~au falls, particles with 
$r <$ 40~\mum\ drift outward more rapidly. The maximum inward drift time 
for particles with $r \approx$ 100~\mum\ to 10--100~cm is approximately 
independent of the gas surface density.
}
\end{center}
\end{figure}

Figure~\ref{fig:gasdrag} shows the drift time for solids of different sizes at 1\,au in gaseous reservoirs with a range of masses. 
{Larger solids drift inward toward the star under gas drag (solid symbols) while radiation pressure drives small, weakly coupled particles outward, where they become colder (open symbols).}
For solids that produce an IR excess (size $< 300\,\mum$), the drift times are short ($< 10^2-10^4$\,yr) for gaseous reservoirs at the level $10^{-6}-10^{-5}$ of MMSN \citep[e.g.,][]{weiden1977a,raf2004,knb2016}.
%
%
Thus, for the amount of gas liberated in the first collision in the example above
($1.5\times 10^{-3}\,\mearth \sim 1.5\times 10^{-6}$ MMSN), 
the particles that produce an IR excess would have a drift time $\lesssim 4000$ yr. 
If all of the gas liberated in each impact survives until the end of the giant impact phase, the cumulative reservoir  
(of $7\times 10^{-3}\,\mearth$ or $\sim 0.7\times 10^{-5}$ MMSN)  
would take $\sim 100$ years to remove the debris signature from the final impact. 
In either case, the accompanying debris signature would be removed on a time scale that is a small fraction of the lifetime of the giant impact phase ($\sim 100$ Myr).

{
Although a rigorous assessment of the removal time requires more detailed calculations
of the long-term evolution of gas liberated from planets and the associated radial drift 
from gas drag and radiation pressure,}
these estimates assume a fairly conservative volatile fraction of 1\% by mass for the 
planet-building material. Carbonaeceous chondrites have volatile fractions that 
are as much as $\sim 10$ times larger \citep[e.g.,][]{alexander2022, piani2020}. 
If the planets and protoplanets form from more volatile rich material, the regenerated 
gas mass is correspondingly larger.
{
In our model, more massive regenerated gas disks have longer radial drift times 
(Fig.~\ref{fig:gasdrag}); however, the drift times are still very short compared 
to the $\sim$ 100~Myr lifetime of the giant impact phase.
}


\section{Gas Delivery by Comets and Asteroids}
\label{sec:comets}

In contrast to the above self-reliant 
pathway, in which
giant planet impacts 
contribute the volatiles needed to remove their own debris signatures, volatiles may also be delivered to the terrestrial zone by other mechanisms \citep[e.g.,][]{greenberg1989,morbidelli2000,vokrouh2000,
raymond2004,obrien2014}. 
For example, volatile-rich planetesimals
scattered from beyond several au into the terrestrial zone may outgas and/or evaporate, and occasionally bombard terrestrial planets. While this mechanism is often invoked to transport water and organic molecules to terrestrial planet surfaces, the difficulty of hitting a planet implies that if this mechanism is
successful, 
a much larger reservoir of volatiles is delivered to replenish a gaseous inner disk. 
\medskip

{\bf Comets.}\quad Taking the Solar System as a guide, we consider
a specific example scenario in which comets deliver water to Earth. 
The mass in comets that impact Earth is
\begin{equation}
    M_{\rm c,imp}= M_{\rm c}\,N_{\rm orb}\,P_{\rm c},
    \label{ImpactMass}
\end{equation}
where $M_{\rm c}$ is the total mass in comets sent to the inner Solar System, $P_{\rm c}$ is the probability that the comet will hit Earth on a single orbital passage, and $N_{\rm orb}$ is the number of times the comet orbits the inner Solar System before it is destroyed or scattered away. 

The collision probability $P_{\rm c}$ is very small. As described by 
\citet{sinclair2020}, comets entering the inner Solar System will generally have velocities $ \sqrt{2} v_{\rm orb}$ when they reach 1~au, where $v_{\rm orb}$ is the orbital velocity of the Earth
\citep[see also][]{schlichting2012a,obrien2018,brunini2018,martin2021}. 
The velocity of the comet relative to Earth is $v_{\rm rel} \simeq 0.4-1.7\,v_{\rm orb}$ depending on whether the comet encounters Earth at pericenter or far from pericenter. These relative velocities correspond to an effective area for capture of $A_{\rm eff} = A_\oplus [1 + (v_{\rm esc}/v_{\rm rel})^2]$ where $A_\oplus$ is the physical cross-sectional area of Earth ($\pi R_\oplus^2$), and $v_{\rm esc}$ is its escape speed. Because Earth's escape speed is 11\,\kms, there is no significant gravitational focusing for comets, given their typical relative velocity of 
$\sqrt{2} v_{\rm orb}.$
The Earth's effective area is tiny compared to the size of the inner Solar System, resulting in a collision probability of
$P_{\rm c} = A_{\rm eff}/\pi(1\,{\rm au})^2 \simeq 
(R_\oplus/1\,{\rm au})^2 = 2\times 10^{-9}$.

As a result, if comets deliver some fraction of an ocean of water to Earth, they also deliver a much larger mass of volatiles to the inner Solar System. 
We can estimate the total mass in volatiles delivered to the inner Solar System 
using equation~\ref{ImpactMass}, 
the above estimate for $P_{\rm c}$, 
and an estimate for $M_{\rm c,imp}$ from the literature. 
%
%
\citet{sinclair2020} estimate that comets deliver $\sim 10^{-4}$ oceans to Earth, equivalent to a total mass 
$\sim 2\times 10^{-8} f_{\rm vol}^{-1}\,\mearth$ if comets have a volatile fraction of $f_{\rm vol}$. This estimate implies 
{that the combination}
\begin{equation}
    M_{\rm c}\,N_{\rm orb} = 
        M_{\rm c,imp}/P_{\rm c} =
        10\,\mearth\,f_{\rm vol}^{-1},
    \label{OrbMassComets}
\end{equation}
or 
$M_{\rm c}\,N_{\rm orb}= 20\,\mearth$ if $f_{\rm vol} = 0.5.$ That is, comets with a total mass $M_{\rm c}$ orbit within the inner Solar System an average of $N_{\rm orb}$ times each, 
which is equivalent to 
{an effective mass of} $M_{\rm c}\,N_{\rm orb}=20\,\mearth$ of comets each making one pass through the inner Solar System. 



We can estimate the volatiles lost by the orbiting bodies
using Solar System comets as a guide. 
If each comet has a mass $m_{\rm c}$ (in gas and dust) and outgasses $\Delta m_{\rm g}$ of volatiles per passage, the effective mass of $M_{\rm c}\,N_{\rm orb}$ corresponds to $M_{\rm c}\,N_{\rm orb}/m_{\rm c}$ comet passages that release a total of $M_{\rm c}\,N_{\rm orb}\,\Delta m_{\rm g}/m_{\rm c}$ in volatiles.
If each comet has the mass of Comet Halley ($\sim 2\times 10^{17}$g in gas and dust)
and
outgasses $2\times 10^{14}\,$g of volatiles 
in one passage
\citep[as estimated for comet Halley's appearance in 1910;][]{hughes1985},
%
the orbiting comets would release a total of 0.02\,\mearth\ in volatiles 
(i.e., $\sim 86$ Earth oceans
$\simeq 2 \times 10^{-5}$ MMSN) 
into the inner Solar System 
\citep[see also][]{thomas1991,jewitt1999,lisse2002}.



{Similarly, we can estimate $M_c$, the total mass in comets sent to the inner Solar System, by estimating $N_{\rm orb}$ from a consideration of}
the lifetimes of the Jupiter Family Comets (JFCs). 
The orbital properties of JFCs today suggest they have active lifetimes of 12,000\,yr 
\citep{levison1997}. For a typical JFC orbital period of 6\,yr, JFCs experience $\sim 2000$ returns to the inner Solar System 
\citep[][and references therein]{weissman2020}. Since comets lose 5--20\,m of surface volatiles per orbit \citep[e.g.,][]{huebner1986,marboeuf2016}
and a Halley-like comet will lose 50\% of its mass in the first 20\% of these passes 
\citep{hughes1985}, comets should strike the Earth within the first few 100 orbits to deliver most of their mass to Earth. 
If $N_{\rm orb} = 400$, $M_{\rm c} = 0.05\,\mearth$, a small fraction of the original reservoir of 10--20\,\mearth\ from which JFC comets originate. 

These orbiting bodies can live as active comets for $\sim 2000$ orbits, evaporating and shrinking away. If all of the $M_c = 0.05\,\mearth$ evaporates and $f_{\rm vol}=0.5$, the comets deliver $0.025\,\mearth$ in volatiles ($\sim 2.5 \times 10^{-5}$ MMSN)
to the inner Solar System, 
comparable to the mass needed to remove the dusty debris signature of terrestrial planet formation (Figure~\ref{fig:mgas}, dotted blue line). 
If comets contribute more than $10^{-4}$ of Earth’s oceans, they would deliver even more gas into the inner Solar System. If some fraction of comets are ejected before they can fully dispense their volatiles, comets would contribute less gas.

\medskip

{\bf Asteroids.}\quad In lieu of comets as the source of Earth's water, \citet{morbidelli2000} estimate that the asteroid belt beyond 2.5\,au could have delivered more than an ocean of water to Earth \citep[see also][]{vokrouh2000}.  
In their simulation, 
which assumes an initial reservoir of $M_{\rm ast} \sim 4$\,\mearth\ in asteroids located beyond 2.5\,au, 
the formation of Jupiter disrupts the asteroids, sending some fraction inward to smaller radii. Each asteroid has a probability of $1.3 \times 10^{-3}$ of colliding with Earth. As a result, $5\times 10^{-3}\,\mearth$ of asteroids impact Earth, delivering $5\times 10^{-4}\,\mearth$ of water, if the asteroids have the volatile fraction of carbonaceous chondrites (10\%). 

The excited asteroids would deliver much more water to the inner Solar System, as can be inferred from the reported results. Figure 1 of 
\citet{morbidelli2000} indicates that a fraction $f_{\rm in} \sim $~10--15\% of all asteroids are deflected onto orbits with perihelia $\lesssim 1$\,au. If the asteroids maintain these orbits long enough to completely outgas their volatiles, they would deliver 
a total mass of $f_{\rm in}\,f_{\rm vol}\,M_{\rm ast}\sim 0.04-0.06\,\mearth$ in volatiles to small radii, more than that needed to remove dusty debris signatures (Figure~\ref{fig:mgas}, dotted red line). Their contribution to the gas budget would be reduced if the asteroids are ejected dynamically before they can evaporate or if they become inactive asteroids before releasing all of their volatiles.

\medskip

{\bf Timing.}\quad
While both comets and asteroids can deliver a significant mass of gas to the inner Solar System, the timing of the delivery depends on the dynamical instabilities that trigger their journey. 
For comets, the gas is likely to have been delivered at about the right time to sweep away the debris signature of Earth's formation. In the Nice model, the dynamical instability involving the giant planets that led to the bombardment of the inner Solar System began sometime in the first 100 Myr \citep{nesvorny2018}, similar to the timescale for the Moon-forming impact, i.e., the last major impact in forming the Earth \citep{canup2021}. 

For asteroids, the situation depends on how the scattering occurs. If the formation of a gas giant like Jupiter launches the asteroids 
\citep[e.g.,][]{morbidelli2000,vokrouh2000}, the gas is likely to be delivered too early to be helpful in erasing the signature of terrestrial planet formation. Gas giant formation likely occurs at $< 10$\,Myr, when a primordial gas disk is present, much earlier than the Moon-forming impact is thought to have occurred. The timing would work out better with asteroid belts that are instead destabilized by the formation of lower mass ice giants (e.g., Neptunes) that form and migrate on a longer timescale. 
The time scale for clearing an orbit is related to the local orbital period, i.e., 
$t_{\rm clear} \sim 10^9 ~ P ~ (M_P / \mearth)^{-2}$ \citep{tremaine1993, margot2015}, 
or 30~Myr for a Neptune-mass planet at 4 au.


\medskip


{\bf Applicability Beyond the Solar System.}\quad 
While the above examples are based on scenarios for our own Solar System, similar 
events may have occurred in the histories of other planetary systems, given the known properties of protoplanetary disks and mature exoplanets. Firstly, the initial cometary reservoirs of other stars are likely to be similar.
At $\sim 1$\,Myr, T Tauri stars are surrounded by dust disks 10s of au in size with median dust masses $\sim 6\times 10^{-5}$ of their stellar mass
\citep[e.g.,][]{villenave2021, manara2022} or $\sim$ 20\,\mearth\ for a $1\,M_\odot$ star. 
Both theory and the demographics of debris disks suggest that protoplanetary disks form planetesimals with modest efficiency $\sim 10\%-50$\% \citep[e.g.,][]{simon2016,li2018,abod2019,carrera2020,klahr2020,
gole2020,rucska2021,nkb2022}. As a result, it seems quite plausible that a large fraction of protoplanetary disks have reservoirs of $\gtrsim 10\,\mearth$ in planetesimals at distances of tens of au.
If giant planets are also present in these disks and undergo migration, a small fraction of these solids may eventually be delivered into the inner, terrestrial planet regions of the system.


Indeed, a large fraction of stars like the Sun are known to harbor giant planets at outer Solar System distances. 
Microlensing studies imply 
an 
average of 0.8 planets per star for planets  $>8\,\mearth$ at orbital distances 1--15\,au \citep{suzuki2016,poleski2021}.
Microlensing observations 
also point to a large population of planets in the 5--60 \mearth\ mass range that either orbit stars at Neptune-like or Uranus-like distances (i.e., beyond $\sim 10$ au) or are free-floating, with about 3 times as many 10\,\mearth\ planets beyond 10\,au as at 0.1--10\,au \citep{poleski2021,gould2022b}.
The above results are consistent with all stars having 1--2 Neptune-mass planets at 
5--30 au. 
Because the escape velocity of a Neptune is 20--25~\kms\ and the orbital velocity at 10~au is 13~\kms, a Neptune-mass planet is sufficient to scatter icy planetesimals into the inner planetary system. Because all stars have one or more Neptunes, this mechanism for delivering volatiles to the inner Solar System could be generic. 


Models of the growth and migration of giant planets find that as they rapidly grow in mass (and migrate) during the rapid gas accretion phase, they destabilize the orbits of nearby planetesimals, which are scattered into eccentric orbits that deliver water to the inner few au of planetary systems \citep{raymond2017,ronnet2018}. 
Dynamical interactions between multiple (neighboring) giant planets would also deliver volatiles to inner planetary systems. 
The broad distribution of orbital eccentricity among the known exoplanets suggests that such instabilities are common in the histories of other planetary systems \citep{chatterjee2008,ford2008,juric2008}. The much higher eccentricities of exoplanets compared to the Solar System planets imply that they experienced more dramatic instabilities \citep{raymond2010}.
\medskip

Thus, a tenuous gas disk can be acquired via ``takeout'' (i.e., the liberation of planetary atmospheres in giant impacts) or ``delivery'' (i.e., by asteroids and comets launched into the terrestrial planet region) at a sufficient level to remove the dusty debris signature of terrestrial planet formation.
{While larger solids will migrate under gas drag into the star,  smaller solids will migrate to the outer edge of the gas disk ($> 3$\,au; Figure 2), thereby removing the warm excess signature of terrestrial planet formation.}
The strong stellar winds that are expected to accompany young stars offer an additional dust removal pathway via stellar wind drag.




\section{Stellar Wind Drag}
\label{sec:winds}


Throughout the giant impact phase 
(10--100 Myr), the local environment within the
terrestrial zone is severe. Compared to the Sun, young solar-type
stars have large magnetic fields 
\citep[$B_{\star} \lesssim$ 3--5 kG;][]{johnskrull2007,yang2008,
johnstone2014,vidotto2021} 
and short rotational periods 
\citep[$P_{rot} \approx$ 2--8 days;][]{hartmann1986,herbst2001,
herbst2007,johnstone2015,vidotto2021},
which power substantial mass loss rates
\citep[$\dot{M} \sim 1-5 \times 10^{-11}$ \msunyr;][]{cohen2014,
ofionnagain2018,vidotto2021} and
considerable X-ray and EUV luminosities
\citep[$L_X \sim 10^{-4} \lsun$; 
$L_{EUV} \sim 10^{-3} \lsun$;][]{tu2015,johnstone2021,vidotto2021}.
As described by Vidotto 2021, there are few direct observations of stars in the 10--100 Myr age, although multiple lines of evidence point to the possibility of large wind mass loss rates, with a range of $\sim 10^{-10}-10^{-12}\,\msunyr$.





To estimate the importance of the stellar wind on small solids during the
giant impact phase, we adopt typical values for young solar-type stars,
a mass loss rate $\dot{M} \sim 1.5 \times 10^{-11} \msunyr$
($10^{15}$~\gs) and a wind velocity $v_w \approx$ 500~\kms\
\citep[see also][]{chen2005a,chen2005b,chen2006}. Compared to 
the current Sun, the wind velocity is the same, but the mass loss rate 
is 1000 times larger.
In the absence of gas within a circumstellar disk, the stellar wind 
induces radial drift of small particles \citep[e.g.,][]{burns1979}.
In a frame of reference moving with an orbiting particle, radiation 
and the wind from the central star oppose the particle's motion,
dragging the particle inward. For a central star with $L_\star \approx$
1 \lsun, the time scale for radiation (Poynting-Robertson; PR) drag on
particles with bulk density $\rho_s$ = 3~\gcmc\ is 
\begin{equation}
\label{eq:prdrag}
t_{\rm PR} \approx 20
\left ( \frac{a}{\rm 1~au} \right )^2
\left ( \frac{r}{\rm 1~cm} \right ) ~ {\rm Myr} ~ .
\end{equation}
For particles smaller than 1--2~\mum, radiation pressure dominates 
PR drag and ejects particles from the planetary system on the local 
dynamical time. 

For 1~\mum\ and larger particles orbiting the Sun, wind drag is a factor 
of $\sim$ 5 times smaller than PR drag, $t_w \sim 5 t_{\rm PR}$ 
\citep{burns1979}. However, 
for a young star with a factor of 100--1000 times larger mass loss rate, wind drag is a factor of 100-1000 times more effective than for the Sun \citep[e.g., see also][]{plavchan2005,chen2005a,chen2005b,
chen2006,spalding2018,spalding2020}. 
For a 1~cm particle orbiting at 1 au around a 
10~Myr-old solar-type star with a mass loss rate of $10^{15}$~\gs, the 
time scale for wind drag is then $t_w \sim 10^5$~yr, and the wind drags 
micron-sized particles into the central star on a decade time scale. 

In a somewhat different approach from \citet{burns1979}, 
\citet{spalding2018} and \citet{spalding2020} consider radial drift 
through a magnetic stellar wind from a rapidly-rotating young star
\citep[see also][]{plavchan2005,plavchan2009,klacka2013}. 
The radial drift time scale is similar:
\begin{equation}
\label{eq:drift}
\tau_w \approx 1.25 \times 10^8 \left ( \frac{r}{\rm 1~cm} \right )
\left ( \frac{a}{\rm 1~au} \right )^2
\left ( \frac{\dot{M}}{10^{12} ~ \gs} \right )^{-1} ~ {\rm yr}.
\end{equation}
For a typical mass loss rate throughout the giant impact phase,
$\dot{M} \approx 10^{15}$~\gs, the drift time for a 1~cm particle
at 1~au is then similar to the $\sim 10^5$~yr estimate from the 
relations in \citet{burns1979}. 


For the wind to be an effective removal mechanism, particles must drift
radially inward more rapidly than they collide with other particles.
Within a swarm of solids undergoing a 
collisional cascade at 1~au during the giant impact phase, gravitational 
interactions and physical collisions deflect solids and spread them 
radially inward and 
outward \citep[e.g.,][]{gold1978,gold1982,horn1985,shu1985,stewart2000}. 
Among particles with radius $r$, physical collisions with other particles 
of radius $r^\prime \gtrsim r$ completely destroy the particle; 
collisions with much smaller particles yield a modest amount of debris
\citep[e.g.,][]{wyatt2007a,wyatt2008,kb2016a,kb2017a}. 
The time scale for physical collisions to grind solids to dust that can 
be removed by radiation pressure is typically smaller than the time scale 
for radial expansion. Thus, effective wind removal requires $t_w \lesssim 
t_c$, where $t_c$ is the collision time.

To estimate an appropriate $t_c$, we consider the rate of physical 
collisions between a particle with radius $r$ and all other 
particles\footnote{Although smaller 
particles can catastrophically fragment particles with radius 
$r$, including these collisions decreases the removal time by
an amount, $\sim$ 10\% to 50\%, that depends on the bulk 
properties of the solids and the collision velocities. To avoid
additional free parameters, we avoid this complication for this
first exploration of relative removal times.}
with radii $\gtrsim$ $r$. 
For any particle in the swarm, the collision rate with all other solids
is $dn/dt = n \sigma v$, where n is the number density, $\sigma$ is the 
physical cross-section, and $v$ is the relative velocity of the solids.
For collisions between a particle with radius $r$ and other particles with
radius $r^\prime \ge r$, we write $n \sigma = N(\ge r) \sigma / V$, where 
$N(\ge r)$ is the total number of particles with radius $\ge$ $r$ and 
$V$ is the volume. We substitute $A_d (\ge r)$ for $N \sigma$ and 
$V = 4 \pi a^2 H$ for an annulus of solids with width $a$ and vertical 
scale height $H$ at a distance $a$ from the central star. As additional 
approximations, $H = \imath a$, $v = e v_K = e a \Omega$, and 
$e = 2 \imath$, where $e$ ($\imath$) is orbital eccentricity
(inclination) and $\Omega$ ($P = 2 \pi / \Omega$) is the angular frequency
(orbital period). Finally, we
adopt a standard power-law size distribution, $n(r) \propto r^{-3.5}$ for
small particles. The required surface area of the swarm is then
$A_d(\ge r) = A_d (1 \mum / r)^{1/2}$, where we assume that the smallest
particle in the swarm has $r \approx$ 1~\mum, consistent with radiation
pressure removing smaller particles. Setting $A_d = 4 \pi a^2 \ldlstar$
yields a simple expression for the collision rate,
$dn/dt \approx (4 \pi/ P) (1\mum\ / r)^{1/2} \ldlstar$. Defining the
collision time as $t_c = (dn/dt)^{-1}$, 
\begin{equation}
\label{eq: t-coll}
t_c = \left ( \frac{P}{4 \pi} \right ) ~ \left ( \frac{r}{1~\mum} \right)^{1/2} ~ \left ( \frac{L_\star}{L_d} \right) ~ \approx ~ 
10^5 \left ( \frac{P}{\rm 1~yr} \right ) ~ \left ( \frac{r}{\rm 1~cm} \right )^{1/2} \left ( \frac{10^{-4}}{\ldlstar} \right ) ~ {\rm yr} ~ .
\end{equation}

\begin{figure}[t]
\begin{center}
\includegraphics[width=6.5in]{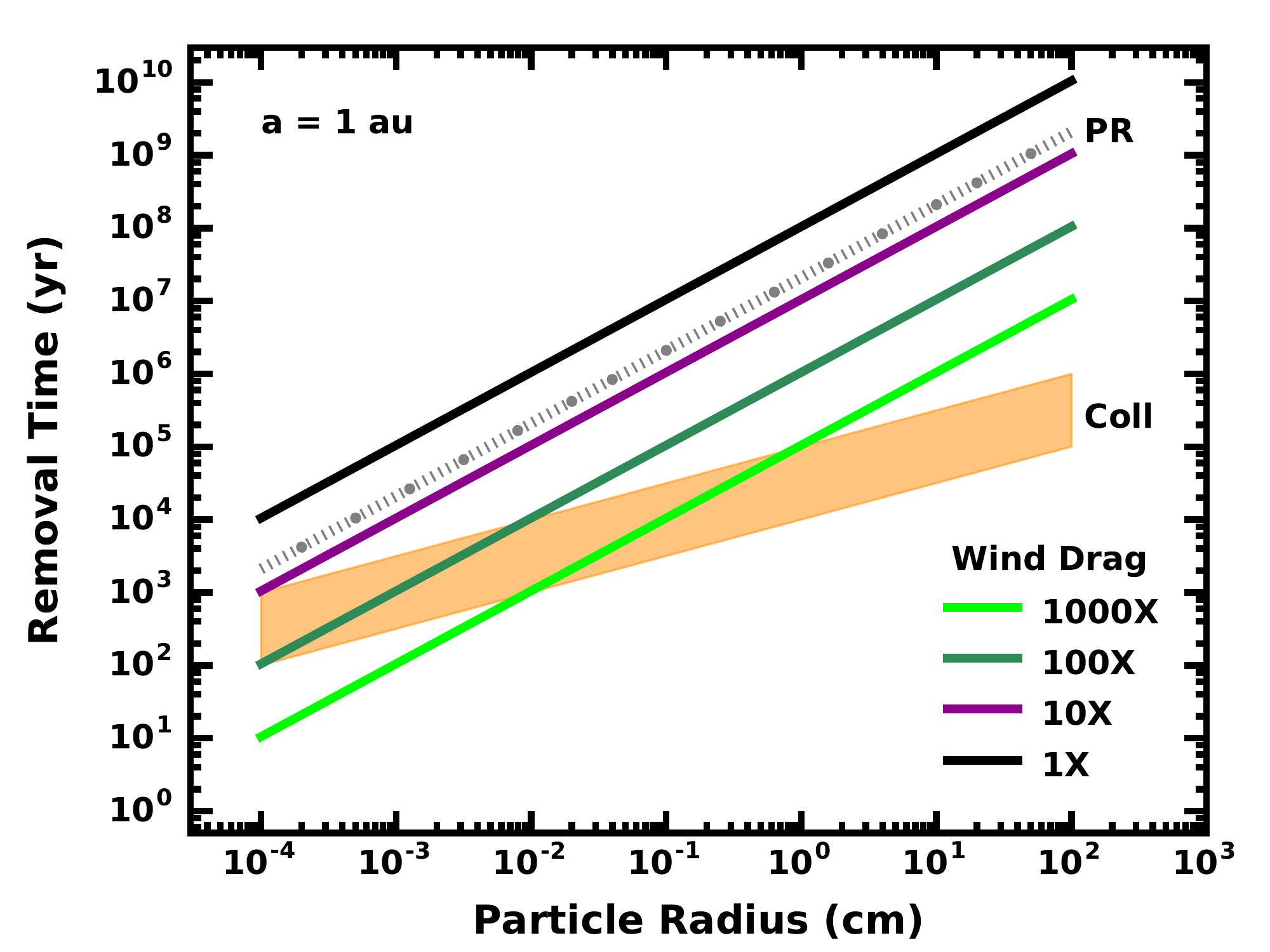}
\vskip -2ex
\caption{\label{fig: drift}
Comparison of radial drift time scales for PR drag (dotted black line)
and for stellar wind drag (black, purple, dark green, and light green 
solid lines) with the collision time for \ldlstar\ = $10^{-4} - 10^{-3}$
(orange band).
The legend indicates the stellar mass loss rate in units of $10^{12}$~\gs.
}
\end{center}
\end{figure}

Fig.~\ref{fig: drift} illustrates the variation of $t_w$ and $t_c$ with
particle radius at 1~au for \ldlstar\ = $10^{-4} -10^{-3}$ (orange band)
and a range of stellar mass loss rates, 1--1000 times the current mass 
loss rate of the Sun (black, purple, dark green, and light green lines). 
As a  reference, the dotted line is the PR drag time scale. Collision 
times for small particles in a system with $\ldlstar\ \gtrsim 10^{-4}$ 
are much shorter than $t_{\rm PR}$. Poynting-Robertson drag is ineffective at 
removing small particles from a collisional cascade around a solar-type 
star with \ldlstar $\gtrsim 10^{-4}$ \citep[e.g.,][see below]{wyatt2011}. 
For the current Sun, the time scale for solar wind drag is roughly five 
times longer than $t_{\rm PR}$ \citep[e.g., Fig. 11 of][]{burns1979}. Thus, a 
wind with the current Sun's mass loss rate cannot remove small particles 
from a collisional cascade. When the central star has a higher mass loss 
rate, $\sim 10^{13}$~\gs, the wind is more effective at removing small 
particles than PR drag. However, the collision time remains shorter than 
the drag time for all particle sizes. As the stellar mass loss rate 
increases to $10^{14}$~\gs, $t_w$ becomes comparable to $t_c$ for small 
particles, $r \lesssim$ 1--10~\mum. For the largest stellar mass loss 
rates, $\sim 10^{15}$~\gs, the drift time scale for 1--100~\mum\ 
particles is shorter than the collision time. At these high mass loss 
rates, the smallest particles drift through the wind before colliding 
with any other similarly-sized particle in the swarm.



{
If collisional cascades in the terrestrial zone generate wavy size distributions
\citep[e.g.,][]{campo1994a,obrien2003,wyatt2011,kb2016a,kb2020}, then the orange 
band in Fig.~\ref{fig: drift} would have a wavy shape, with minima in the removal 
time where the size distribution has maxima. For rocky solids with the bulk properties
of basalt \citep[e.g.,][]{benz1999}, the wave amplitude is a factor of $\sim$ 10 
at 1-100~\mum\ and damps quickly at larger radii \citep[e.g., Figs. 4--5 of][]{kb2016a}. 
In Fig.~\ref{fig: drift}, the removal time for collisions would then be a factor of
$\sim$ 10 smaller at 1~\mum, a factor of $\sim$ 10 larger at 10~\mum, and roughly
the same at $\gtrsim$ 100~\mum. Relative to the example in Fig.~\ref{fig: drift} with a 
very powerful stellar wind (1000 times the solar wind), collisions would compete
with wind-induced drift for 1--3~\mum\ particles. However, the wind would dominate 
removal at 3--100~\mum. Deriving the size distribution of a collisional cascade in
the presence of a powerful stellar wind requires a more detailed calculation that is
beyond the scope of the present effort.
}

To estimate the impact of wind drag on the IR excess, we consider a simple
model for the removal of 1-100~\mum\ particles.
{For simplicity, we ignore the effect of a wavy size distribution.}
Without solar wind removal
and in an equilibrium collisional cascade, the production and removal 
rates of small particles from collisions is 
$\dot{M}_d = N m / t_c$ = constant \citep[e.g.,][]{wyatt2011,kb2016a}, where $N$ ($m$) is the number (mass) of particles with radius $r$.
We adopt a wind removal rate, $\dot{M}_{wd} = N m / t_w$. 
Combining the two relations, $\dot{M}_{wd} \approx \dot{M}_d ~ t_c / t_w$.
In a system with $t_w < t_c$, the wind removes solids faster than the 
collisional cascade. 

In Fig.~\ref{fig: drift} with $t_c \approx t_w$ at $r \approx$ 100~\mum\ 
for \ldlstar\ $\sim 10^{-3}$ and a wind with $\dot{M} = 10^{15}$\gs,
$\dot{M}_{wd} \approx \dot{M}_d (r / 100~\mum)^{-1/2}$; at 1~\mum\ (1 cm), 
the wind mass loss rate is 10 (0.1) times the steady-state mass loss rate 
from the collisional cascade. With these mass loss rates from wind 
removal, the number density of particles at 1~\mum, 10~\mum, and 100~\mum\ 
is 0.10, 0.33, and 0.5 times the number density in the equilibrium cascade.
The total cross-sectional area of small particles is then a factor of 
$\sim$ 5 smaller. Using the same approach, the reduction in IR excess for 
a system with \ldlstar\ $\sim 10^{-4}$ is a factor of $\sim$ 50. In a 
solar-type star with a large mass loss rate, radial drift from the 
stellar wind significantly reduces the IR excess from a collisional 
cascade.

While this analysis demonstrates the effectiveness of radial drift in
removing an IR excess at 1~au, it is important for collisions to dominate 
at $a \gtrsim$ 10--20~au where $\sim$ 25\% of solar-type stars are known to have bright
IR excesses \citep[e.g.,][and references therein]{carp2009a,bryden2009,
sibthorpe2018}. If solids at $a \gtrsim$ 10--20~au have the same physical
properties as those at 1~au, the drift time scale, $t_w \propto a^2$,
grows more rapidly than the collision time scale, $t_c \propto a^{3/2}$.
Extrapolating Fig.~\ref{fig: drift} to 30--40~au, the time scales for
drift and collisions are roughly equal for particles sizes 3--5~\mum\
when \ldlstar\ $\sim 10^{-3}$ and $\dot{M}_w \sim 10^{15}$~\gs.

At large $a$, the lower bulk densities of icy particles yield shorter 
time scales for both removal mechanisms: $t_w \propto \rho_s$ and 
$t_c \propto \rho_s$. For the solar wind, observations suggest a roughly
constant wind speed and a particle density $n \propto a^{-2}$ 
\citep[e.g.,][]{bagenal2016,khabarova2018}. Compared to 1~au, the drift 
time then varies only with $a$ and $\rho_s$: $t_w \propto \rho_s a^2$. 

In a collisional cascade, two factors modify $t_c$ at large $a$:
(i) the sizes of the largest objects involved in the cascade \rmax\ and
(ii) the ratio of the collision energy to the binding energy, which is
encoded in $\alpha = M_d / (\dot{M}_d ~ t_c)$ \citep{kb2016a}. At 
30--40~au, collision velocities (energies) are a factor of $\sim$ 6 (36)
smaller than at 1\,au
for similar orbital eccentricities. Binding energies of small 
solids are a factor of 20--300 smaller \citep[e.g.,][]{lein2012,
schlicht2013}, which results in $\alpha_{\rm 35~au}$ $\sim$ 
0.25--1.1 $\alpha_{\rm 1~au}$ \citep[see Fig.~2 of][]{kb2017a}. With smaller
collision energies, large solids that were collisionally disrupted at 1~au
are `safe'; only somewhat smaller particles take part in the collisional 
cascade, and \rmax\ $\sim$ 30--100~km instead of $\gtrsim$ 300~km. With
$t_c \propto \alpha \rmax$, time scales for a collisional cascade with icy
particles at 30--40~au are then a factor of $\sim$ 3--50 times shorter 
than those with rocky particles. Combined with the variation of $t_c$ with
$a$, the collision time scales for 1~\mum\ and larger icy particles at 
30--40~au are always shorter than the time scales for radial drift within
a stellar wind, even at the largest mass loss rates. Thus, if radial 
drift is important at 1~au, it should have little impact on small solids
at much larger $a$, consistent with the more common appearance of IR excesses at larger orbital distances.

\section{Discussion}
\label{sec:disc}


As described in the previous sections, the warm dusty debris signature that is expected to accompany the formation of terrestrial planets may not be robust. Processes that are central to our current picture of the assembly and evolution of terrestrial planets---early planetary atmospheres and giant impacts (Section \ref{sec:atmospheres}), water delivery by asteroids and comets (Section~\ref{sec:comets})---can potentially regenerate a tenuous gaseous disk that removes the dusty debris via gas drag. 
In addition, drag from the strong stellar winds expected from young stars (10-100 Myr old) can also efficiently remove dusty debris (Section \ref{sec:winds}). 

These processes may help account for the puzzling rarity of warm debris signatures at the expected epoch of terrestrial planet formation, despite the $\sim 20$\% inferred occurrence rate of terrestrial planets orbiting mature solar-type stars. 
{
Because the strengths of gas drag and stellar wind drag are expected to vary from system to system---stellar wind drag depends on the wind mass loss rate, which can vary widely, and gas drag depends on how much of a gas disk is regenerated---the small fraction of systems with a detectable excess may simply reflect the fraction in which these mechanisms do not operate efficiently, i.e., they have weak stellar winds or did not regenerate a significant gas disk.}





\subsection{Photoevaporative Disk Wind}
\label{sec:diskwind}
While our simple estimates point to the potential impact of gas drag and stellar wind drag on debris signatures, understanding their true impact requires detailed modeling that takes into account additional processes. In particular, a photoevaporative disk wind driven by 
stellar X-ray and UV (XUV) irradiation  \citep[e.g.,][]{ercolano2017,nakatani2018a,nakatani2018b}
may deplete a regenerated gas disk, reducing its effectiveness. 
Developed to understand the evolution of much younger T Tauri stars surrounded by high-mass hydrogen-rich disks, the theories
generally predict winds driven 
from distances beyond a few au. Irradiation heats the disk surface to a temperature $T_g$, and the atmosphere becomes unbound beyond the gravitational radius
\begin{equation}
R_g = {G M_*\over c_s^2} \approx 10\,{\rm au} \left(T_g \over 10^4\, {\rm K}\right)^{-1} 
\,\mu\,f,
\label{eq:gravitationalradius}
\end{equation}
where $c_s$ is the sound speed, and the mean particle weight $\bar m = \mu m_H,$ where $m_H$ is the mass of
hydrogen, 
and the correction factor $f$ is  $\sim 0.2$ 
\citep{liffman2003}. 
Thus, gas in the terrestrial planet region is more tightly bound, and more difficult to remove photoevaporatively, than gas at larger distances.
{In addition, both high gas temperatures ($\sim 10^4$\,K) and low mean particle weights ($\mu \sim 1$) are needed to remove gas from the terrestrial planet region. Such high temperatures typically require EUV irradiation, in contrast to the $\sim 1000$\,K temperatures more typical of FUV irradiated disks \citep[e.g.,][]{gortiDH2009}.
Low mean particle weights would be characteristic of  hydrogen-rich gas but not gas rich in water vapor.} 


The impact of disk winds on a regenerated gas disk is difficult to estimate, because empirical wind mass loss rates are poorly known for stars of any age, and the predictions of disk wind theories 
are generally unverified observationally. 
In addition,
because they depend on XUV fluxes, photoevaporative disk wind mass loss rates are also likely to vary greatly even among stars of similar age.
Approximately solar-mass stars in the age range 10-100 Myr are estimated to have XUV fluxes that 
range from T Tauri luminosities to values $\sim 100$ times lower ($L_X = 3\times 10^{28}-3\times10^{30}$ erg\,s$^{-1}$), a consequence of both age and differing initial stellar rotation rates 
\citep{tu2015}.
%


Nevertheless, we can explore a few possibilities. 
For example, in the study by \citet{picogna2019}, 
which explored photoevaporative disk wind mass loss rate as a function of X-ray luminosity,
the total mass loss rate, integrated over all disk radii, is $\sim 10^{-9}\,\msunyr$ for $L_X \sim 10^{29}$\,erg\,s$^{-1}$.
Although most of the mass loss will occur at $R_g$ (eq.~\ref{eq:gravitationalradius}) and beyond,
if $\sim 1$\% of the total mass loss originates within 1\,au, 
it would take $\sim 300$ years to remove 0.001\,\mearth\ of gas from the inner au, a short interval compared to the drift removal time for the dust (Figure~\ref{fig:gasdrag}). If the mass loss rate is 10 times lower, the drift rate is comparable to the gas removal time, and the gas released in giant impacts could plausibly persist long enough to remove the debris signature from a given giant impact. 
%

Importantly, these models assume a hydrogen-rich gas disk, whereas regenerated gas disks may be greatly enhanced in elements heavier than hydrogen, with a corresponding increase in $\mu$ and $R_g$. For example, if the gas disk originates as water vapor that is subsequently dissociated, $\mu=6,$ and 
{$R_g \approx 12$~au if $f=0.2$ and $T_g = 10,000$\,K.} A metal-rich composition will also affect the cooling of the gas, with lower temperatures increasing $R_g$. Larger values of $R_g$ will better shield the terrestrial planet region from photoevaporation 
{and extend the time available to remove the warm excess}.
Models tuned to the conditions in regenerated disks surrounding 10--100 Myr old stars would be very welcome.

Thus, although there are many uncertainties, it seems plausible that photoevaporative disk winds could compromise gas drag as a dust removal mechanism at high $L_X.$
Luckily, if stellar XUV is efficient in reducing the effect of gas drag, the stellar wind associated with the same XUV activity can potentially step in to remove dusty debris via stellar wind drag.
Given the multiple processes that are potentially involved, the range in $L_X$ as a function of age and initial conditions, and the timing of terrestrial planet formation, more detailed calculations are needed to understand which processes dominate under what conditions, and the circumstances under which dust removal, by gas drag and stellar wind drag, is effective.

\subsection{Quick and/or Neat?} 
\label{sec:quick}

The above discussion illustrates how gas drag and stellar wind drag can potentially act to remove the expected dusty signature from giant impacts and 
account for the surprising lack of warm excesses that are expected to characterize late-stage terrestrial planet formation. If processes like these are effective, terrestrial planets may form in the traditional way, via giant impacts, but leave little observational trace of these events (a tidy or ``neat'' planet formation scenario).

An alternative interpretation of the rarity of warm excesses is that terrestrial planets emerge fully formed from the nebular phase and only rarely undergo late-stage giant impacts (a ``quick'' formation scenario). 
Possible motivation for the latter scenario comes from the bimodal size distribution of close in planets (the radius valley at $< 0.3$\, au), which is interpreted as evidence that planets $> 5\,\mearth$ are commonly born with gaseous envelopes that are eroded by stellar irradiation and core-powered mass loss, to a greater or lesser extent, depending on their 
initial envelope mass as well as planet mass and orbital distance 
\citep[e.g.,][]{owen2013,lopez2013,jin2014,owen2016c,ginzburg2016,
owen2017b,ginzburg2018,jin2018,gupta2019,mordasini2020,venturini2020,
rogers2021}. In this picture, multi-\mearth\ planets 
must clearly form quickly (perhaps via pebble accretion), before the dissipation of the protoplanetary disk,
i.e., within the first few Myr. The ability of these planets to reach masses above $1\,\mearth$ quickly at distances $< 1$\,au suggests that $\sim \mearth$ planets forming at $\sim 1$\,au may also 
acquire all of their mass in the nebular phase, emerging fully formed, with no need for late-stage giant impacts \citep[e.g.,][]{mordasini2020,
venturini2020}. 



In the Solar System, evidence for the moon-forming impact
\citep[e.g.,][and references therein]{asphaug2021,canup2021}
and analyses of elemental abundances in Venus,
Earth, and Mars \citep[e.g.,][and references therein]{lammer2021,
lammer2022} provide strong incentive for the role of giant 
impacts in the evolution of at least one planetary system.
However, this evolutionary history may be rare if most planetary systems form their terrestrial planets early, in the nebular phase. 

\subsection{Future Observations} 
\label{sec:future}

Future work on exoplanet demographics and work that explores the plausibility of dust removal mechanisms can help us understand 
whether planetary systems form their terrestrial planets primarily ``quickly'' or ``neatly.''


{\bf Gas reservoirs.} If late-stage giant impacts are a common feature of terrestrial planet formation but the debris they produce is swept away by gas drag, we expect to see tenuous gas reservoirs around  10--100 Myr old stars.
Residual gas disks are generally assumed to be absent based on the lack of detected stellar accretion signatures and {\it in situ} gas detections. But it is difficult to rule out gas at a very low level. Stellar accretion rate measurements are not commonly available below $\sim 10^{-11}-10^{-10}\,\msunyr,$ i.e., below 0.1\%--1\% of the accretion rates typical of classical T Tauri stars \citep[e.g.,][]{riviere2015}. In terms of gas detections, a recent study using the UV transitions of H$_2$ to search for residual gas around $\sim 10$\, Myr old stars detected $10^{-10}-10^{-8}\,\mearth$ of hot H$_2$ ($\sim 1500$\,K) within 1\,au of TWA7 
\citep{flagg2021}. Just the tip of the iceberg, the hot H$_2$ is expected to be accompanied by a much larger mass of cool and/or atomic gas, perhaps a total column density of $\sim 0.1$ g\,cm$^{-2}$, similar to the column density needed for efficient dust removal via gas drag. Further studies along these lines would be of great interest.
Given the low column densities of the replenished gas disks, they are likely to be best probed with atomic diagnostics. If they arise from secondary atmospheres or are delivered by comets and asteroids, they will be enriched in volatiles (e.g., water) rather than dominated by hydrogen, suggesting searches for atomic oxygen line emission.



{\bf Stellar Winds of Young Stars.}
The major challenge in understanding the efficiency of stellar wind drag is 
the current uncertainty in the mass loss rates of young
stars \citep[e.g.,][and references therein]{cranmer2011,
cohen2014,fichtinger2017,ofionnagain2018,gudel2020,vidotto2021}. With few 
direct measures of mass loss rates for 10--100~Myr old stars, 
published estimates are based on observations of older stars (0.1--1~Gyr) 
and models that match the evolution of measured rotation rates and proxies 
of coronal activity. Future direct measurements of mass loss rates for 
stars with ages of 10--100~Myr would 
be extremely valuable. 
Fortunately, indirect evidence for stellar wind drag may be found in 
the combined X-ray/IR demographics of young stars
\citep[e.g.,][]{plavchan2005,chen2005a,chen2005b,chen2009,
plavchan2009,sierchio2010}. Although sample sizes remain modest, some
analyses suggest that young stars with large (small) X-ray luminosities
are less (more) likely to have IR emission from a warm debris disk at 
distances of a $\lesssim$ 2--3~au from the central star, a relation that is expected if stellar winds associated with stellar activity act to remove dust signatures. Augmenting these
studies with more recent X-ray data from {\it Chandra} and the results of
IR surveys with WISE \citep[e.g.,][]{kuchner2016} and Herschel
\citep[e.g.,][]{eiroa2013,sibthorpe2018} might enable a cleaner test 
of a correlation between X-ray emission (a possible proxy for the
stellar mass loss rate) and IR excess emission.

{\bf Terrestrial Planet Atmospheres and Moons.} 
If terrestrial planets with masses 
$\sim$ 1--2~\mearth\ at $\sim 1$\,au form quickly and emerge from
the nebular phase fully assembled, they will have primordial gaseous envelopes with
masses $\lesssim$ 1\% to 10\% of the mass of the planet 
\citep[e.g.,][]{mordasini2012b,mordasini2020}. 
Although various processes
can remove primordial atmospheres with masses $\lesssim$ 0.01~\mearth\
\citep[e.g.,][and references therein]{wyatt2020,kegerreis2020a,
kegerreis2020b,lammer2021}, the most massive envelopes may survive giant
impacts and evaporation. If this dichotomy is borne out by detailed 
calculations, we expect ensembles of 1--2~\mearth\ planets at 1\,au distances to consist of
a mixture of planets with negligible atmospheres and those with puffy atmospheres, 
similar to the situation found for sub-Neptunes and super-Earths at smaller orbital distances.
Future searches for exoplanetary moons may uncover valuable clues as to whether late-stage giant impacts play a significant role in terrestrial planet formation.
If the presence of large moons (like that of Earth) is a signpost of late-stage giant impacts, the incidence rate of moons associated with terrestrial exoplanets can reveal the rarity or commonality of this formation pathway.

\section{Summary and Conclusions}
Giant impacts are a central theme in the story of terrestrial planet origins. They play an important role in our ideas about how terrestrial planets gather their mass and in the expectation that 
a dusty debris signature 
announces the final stages of planetary assembly.
As described here, 
giant impacts also 
figure prominently 
in our ideas about how planets acquire (or lose) their atmospheres and oceans, 
but it is perhaps less appreciated how the loss of planetary atmospheres 
can also regenerate enough of a gas disk to compromise our ability to detect the expected dusty debris signature of terrestrial planet formation.

Similarly, impacts by comets and asteroids are often invoked to account for the origin of Earth's oceans. Because the Earth is a small impact target,  scenarios that deliver appreciable water to Earth would also launch many more non-Earth-impacting bodies into the inner Solar System and 
regenerate a tenuous gaseous disk. 
In either case, a regenerated gas disk, whether achieved by  ``takeout'' (by giant impacts liberating planetary atmospheres) or ``delivery'' (by evaporating comets and asteroids), 
is plausibly hefty enough to whisk away dusty debris via gas drag and erase the anticipated debris signature of terrestrial planet formation.

The regenerated gas disk may be subject to photoevaporation by energetic XUV radiation from the central star, potentially reducing its efficacy in removing the dusty debris. However, the strong stellar wind that will accompany the XUV radiation offers an alternative pathway to remove the dusty debris signature of terrestrial planet formation: stellar wind drag. 
The major uncertainty associated with stellar wind drag is the relatively poorly measured stellar wind mass loss rates at ages of 10-100 Myr. 
%

Thus, 
both gas drag and stellar wind drag provide pathways by which the anticipated debris signature of terrestrial planet formation can be erased. These processes may resolve the conundrum of 
why 
terrestrial planets like Earth and Venus are found to occur relatively commonly around Sun-like stars, but the warm dust debris signatures that are expected to accompany their formation are rarely detected. 

This conundrum might also be resolved in a different way. 
The size distribution of close-in planets may suggest that terrestrial planet formation  departs significantly from the ``classical'' picture, with planets forming completely in the nebular phase and late-stage giant impacts playing little or no role. If this is the dominant pathway for terrestrial planet formation, the primordial disk may obscure the assembly process, making it challenging to detect and study. 
Thus, both the availability of multiple pathways to erase the anticipated debris signature of terrestrial planet formation, as well as the alternative possibility of very early terrestrial planet formation, suggest 
that the observable signposts of terrestrial planets may be more subtle and more challenging to detect than previously believed. 

Exploring these ideas requires 
more detailed models that track the transport of gas between planets, comets, asteroids, and a regenerated gas disk, including important additional processes such as photoevaporation and other gas removal mechanisms. 
Valuable insights can come from new observations, e.g., measurements of stellar wind mass loss rates and confirmation of the presence and longevity of tenuous regenerated gaseous reservoirs. 
Studies along these lines may find evidence for dust removal mechanisms in action.
In parallel, future studies of the mass-size distribution of terrestrial planets and the presence or absence of large moons can also reveal whether terrestrial planets primarily form early, in the presence of a gas disk, and whether giant impacts play a significant role in their formation history.





\begin{acknowledgments}
We thank Hilke Schlichting and the anonymous referee for valuable comments on the manuscript and the Harvard EPS/CfA Planetary Journal Club community for interesting discussions that stimulated some of the ideas considered here. 
JN’s research was supported in part by a Fellowship from the John Simon Guggenheim Memorial Foundation, the Harvard Radcliffe Fellowship Program of the Radcliffe Institute for
Advanced Study at Harvard University, and the Institute for Theory and Computation at the Harvard-Smithsonian Center for Astrophysics. 
JN’s research activities are also supported by the NSF’s NOIRLab, which is managed by the Association of Universities for Research in
Astronomy (AURA) under a cooperative agreement with the National Science Foundation. 
SK's research is supported by the Smithsonian Institution.
\end{acknowledgments}


\bibliography{sfpl}{}
\bibliographystyle{aasjournal}



\end{document}